# A PAXOS based State Machine Replication System for Anomaly Detection

**Manoj Rameshchandra Thakur***
Computer Science Department, VJTI, Mumbai, India
manoj.thakur66@gmail.com
**Sugata Sanyal**
School of Technology and Computer Science,
Tata Institute of Fundamental Research, Mumbai, India
sanyals@gmail.com

**Abstract:** A number of systems in recent times suffer from attacks like DDoS and Ping of Death. Such attacks result in loss of critical system resources and CPU cycles, as these compromised systems behave in an abnormal manner. The effect of such abnormalities is worse in case of compromised systems handling financial transaction, since it leads to severe monetary losses. In this paper we propose a system that uses the Replicated State Machine approach to detect abnormality in system usage. The suggested system is based on PAXOS algorithm, an algorithm for solving the consensus problem in a network of unreliable processors.

## 1. Introduction

Previous approaches towards intrusion and anomaly detection have been based on detecting the deviation of the system from its normal behavior. For instance, a number of anomaly detection techniques inspired by Artificial Immune System have been based on self, non-self-detection. An efficient technique for anomaly detection would be to compare the functioning of the system with replicas of the system serving the same set of requests from the client. Abnormality in such a case would be indicated by a deviation in the functioning of the individual replicas and hence a lack of agreement on the resultant output of the system. In order to manage the consensus problem in our proposed system we use the PAXOS algorithm [1]. The algorithm is generally applied in the circumstances that require durability, wherein the amount of durable states are large. It functions suitably well even in the presence of a limited number of unresponsive replicas. A brief description of the state machine based approach is presented below, details of the system functioning is explained in later sections of the paper.

*State Machine Replication, the approach:* A state machine replication based approach involves replicating a state machine on multiple instances of a system. Each of the state machine replicas begin with the same initial state. When the system receives a client request each of the replicas of the system process the request and based on the output generated, update their individual states. In a normal scenario the resultant state of each of the state machine replicas will be identical. In case of a failure or abnormality, there will be no consensus on the new state of the system. This lack of agreement of the new state of the replicated state machine indicates an abnormality [2].

___________________________________________________

*Corresponding Author

## 1.1. Related Work

A number of approaches have been proposed towards anomaly and intrusion detection. Most of these approaches can be categorized into two categories: Artificial Immune System and Soft Computing. A brief overview of these two approaches is as follows:

Approaches based on Artificial Immune System are inspired by the human immune system. The first lightweight intrusion detection systems based on AIS (Artificial Immune System) was introduced by [13]. 'Danger Theory' based intrusion detection system is presented in [14] [15]. Most of the suggested Intrusion detection Systems which are based on AIS models has used one of the following algorithms: negative selection algorithm, clonal selection algorithm, artificial immune network, danger theory inspired algorithms and dendritic cell algorithms [16].[17] attempts to detect anomaly in electromagnetic signals in a complex electromagnetic environment.[18]focuses on specifically static clonal selection with a negative selection operator.[19] presents an approach to solve the problems with existing intrusion detection systems using autonomous agents. [20] presents an analogy between the human immune system and the intrusion detection system. It also uses genetic operators like selection, cloning, crossover and mutation attempts to evolve the Primary Immune Responses to a Secondary Immune Response. [21] presents a genetic classifier-based intrusion detection system.

The problem of anomaly detection in a system is characterized by lack of exactness and inconsistency. This has encouraged a number of attempts towards intrusion detection based on 'Soft Computing' [22] [23]. 'Soft Computing' techniques attempt to evolve inexact and approximate solutions to the computationally-hard task of detecting abnormal patterns corresponding to an intrusion. [24] applies a combination of protocol analysis and pattern matching approach for intrusion detection.[25] proposes an approach towards intrusion detection by analyzing the system activity for similarity with the normal flow of system activities using classification trees. [26] proposes a Soft Computing based approach towards intrusion detection using a fuzzy rule based system. [27] presents an intrusion detection system based on machine learning techniques.[30] presents a proactive detection and prevention technique for intrusions in a Mobile Ad hoc Networks (MANET).

In spite of the differences in the approach taken by the above mentioned works, each attempts to detect the deviation of the system from its normal functioning. In our suggested approach we use state machine replication technique, often used in fault tolerant distributed systems.

*State Machine Replication*

A number of approaches based on State Machine Replication have been proposed and used in real time systems for fault tolerance [10].[6] explains a general method for implementing a fault-tolerant service by replicating servers and coordinating client interactions with server replicas. [7] explains the implementation of a fault tolerant system using state machine replication. None of the previous approaches for anomaly detection have been exclusively based on state machine replication. The rationale behind using the state machine replication is that this approach is instrumental in failure detection and fault tolerance in distributed systems. And that the problem of failure detection is similar to anomaly detection in real time systems.

The rest of the paper is structured as follows: section 2 explains the system deployment and system components in detail followed by the state machine replication and anomaly detection explanation. Section 4 explains the proposer node election and acceptor node failure handling followed by acceptor node failure handling explanation in section 5. Section 6 addresses the confidentiality and integrity issues related to the system specific packet and section 7 presents the conclusion of the suggested system.

## 2. System Deployment and Components

The suggested system consists of three types of nodes namely [1]:

1) Acceptor node: These are the nodes that represent replicated instances of the application in a network. Acceptor nodes are the ones that actually produce the response to requests initiated by the client. The collection of all the acceptor nodes is referred to an acceptor quorum.
2) Proposer node: Proposer node is the acceptor node that accepts requests from the client and forwards requests to each of the acceptor nodes in the acceptor quorum. It acts as the leader acceptor and ensures consensus among the acceptors by coordinating the *promise* and *accepted* packets from the proposer node.
3) Learner node: Learner nodes are the nodes that upon receiving *accepted* packets send the agreed upon response to the client or take appropriate action in case an anomaly is detected.

The detailed explanation of the *promise* and *accepted* packets is presented in the later part of the paper. The TCP packets used to regulate the suggested PAXOS based system are classified as follows [1]:

**Prepare packets, Promise packets, Accept request packets, Accepted packets**

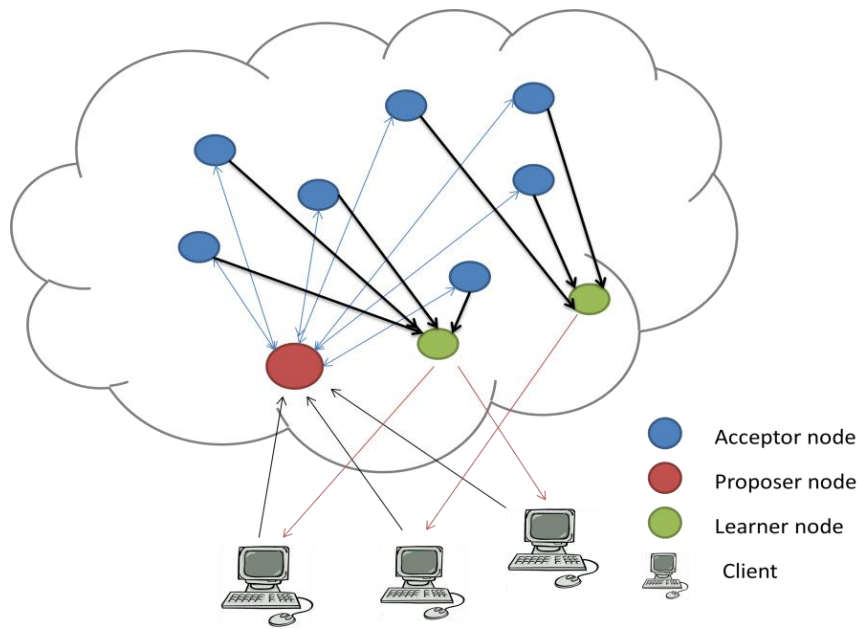

**Figure 1 Network deployment of the proposed system along with the components**

A detailed explanation of the packet types is as follows [1]:

1) Prepare packets
   The Acceptor node on receiving a request from a client creates a prepare packet (the proposal) identified by a number $N$ such that any further prepare packet will have a value $N' > N$. The prepare packet also contains the actual request received from the client. The acceptor node then broadcasts the prepare packets to the acceptor quorum.

2) Promise packet
   The promise packets are initiated by acceptor nodes which are part of an acceptor quorum in response to the prepare packet from the Proposer node. The promise packet sent by an acceptor in response indicates the willingness of the acceptor node to serve the request sent by the client. An acceptor may not respond with a promise packet if it has already served a proposal by the proposer node with a value $N'<N$. The promise packet contains the value $N''$ which is the value of the last successful proposal that it served.

3) Accept request packet
   If the proposer node received the promise packets from majority of the acceptor packets in the acceptor quorum, initiates an accept request packet, signifying that majority of the acceptor nodes have agreed to respond to the client request. If the proposer node does not receive promise packets from majority of the acceptors then it reinitiates the proposal with $N$ as the maximum value received from the promise packets in response.

4) Accepted packets
   Upon receiving the accept request packet from the proposer node, each of the acceptors generate the response to the request from the client and send it to the proposer node and the learner node. The learner, on receiving the response to the client request, sends the response to the client.

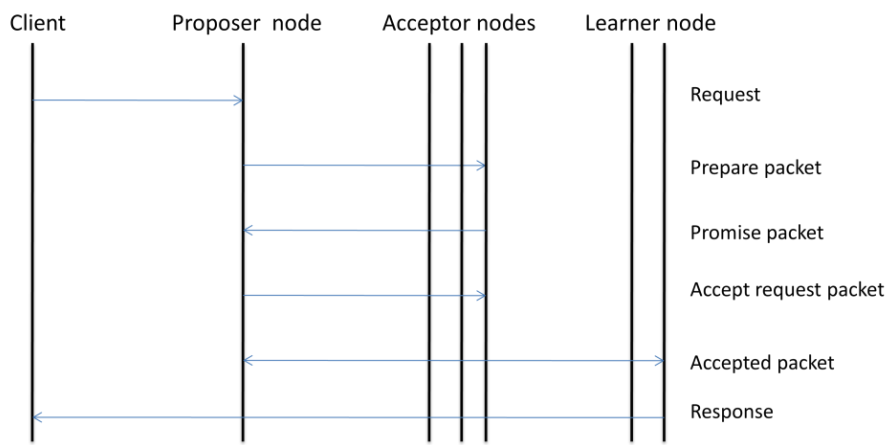

**Figure 2 Flow sequences of system specific packets among the system components**

It must be noted that since the suggested system is based on the PAXOS algorithm [1], the names of the components are based on the names of the components in the PAXOS system.

## 3. State Machine Replication and anomaly detection

Each acceptor node apart from running the application instance, that serves client requests, maintains a state machine which consists of the following parts [28][1]:

- *A set of States*: Each state represents the status of the acceptor node at any given instance. The state machine has a distinguished state called Start state.
- *Inputs*: Inputs represent the client requests to the application.
- *Outputs*: Outputs represent the response generated by the acceptor node in response to the client request. Outputs generated are based on the type of application and the kind of requests that it serves.
- *A transition function (Input x Output x State -> State):* It is the function that determines the next state of the acceptor node given the current state of the acceptor node, the input received in the current state and the output generated in the current state.

In general, real time applications serve complex requests and the responses generated that vary based on the client requests. In such a scenario rather than comparing the inputs received from and outputs generated by the states of the State Machine with discrete values we use Regular Expressions (REGEX)[29] to check inputs and outputs generated for determining the next state of acceptor nodes.

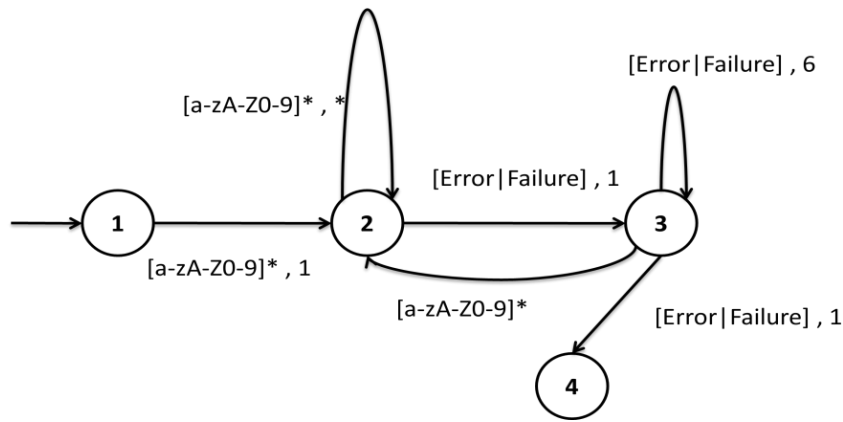

**Figure 3 A typical State Machine**

In the above state machine the labels corresponding to the edges represent:

1) The output generated in a particular state.
2) The numbers of times to consider the transition function for a particular state. For example, if in state 3, the acceptor node generates either an 'Error' or a 'Failure' output (response) six times continuously then the seventh 'Error' or 'Failure' output will cause a state change to state 7. It must be noted that a '*' represents that the state uses the transition function indefinitely.

It must be noted that in the above state machine since each state accepts inputs following the same regular expression, it is not shown in the labels corresponding to the edges.

Each acceptor node after accepting the request from the proposer node performs the execution based on the request and generates the output to be sent in response. Based on the output generated the acceptor node also updates its current state. Once the output and the new state are determined, each acceptor node sends the following tuple to the learner node:

[N, <output generated>, <new state>]

The learner node on receiving all such tuples from all the acceptor nodes in the acceptor quorum, tries to determine the presence of an anomaly in the system by comparing the new statuses sent by each of the acceptor nodes with each other. If the learner node finds a consensus in the new states of the acceptor nodes it returns the output generated as response to the client. However, if the learner node is unable to find a consensus in the new states it detects an anomaly in the system and takes appropriate steps.

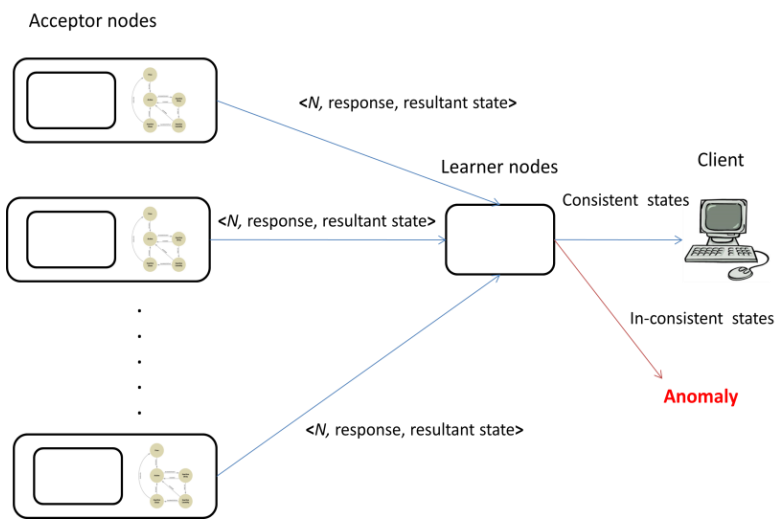

**Figure 4 Anomaly detection by the learner node based on the states sent by individual acceptor nodes**

## 4. Proposer node election and acceptor node failure handling

The proposer node is the central entity regulating the working of the acceptor nodes. It manages 'the promise packets' and sending of 'the prepare packets and the accept request packets'. As a result the acceptor quorum should contain a proposer node at any given instance to regulate the PAXOS based system. To ensure this each acceptor node including the proposer node, are considered as members of a group with the proposer node as the leader. Whenever the leader fails all the remaining healthy members try to offer leadership. Based on some predefined strategy a leader is elected among the contesting members. This newly elected member acts as the new proposer node. The group membership and leader election for the acceptor nodes can be implemented using Apache Zookeeper, a service for distributed co-ordination [3].

## 5. Acceptor Node failure handling

Handling the acceptor node failures and taking appropriate actions based on the failures is crucial in the functioning of the PAXOS algorithm. Consider a scenario wherein there are six acceptor nodes in the system and a proposer node. On receiving a client request the proposer node will initiate prepare packets for each of the acceptor nodes. Based on whether the acceptor nodes can handle the request (depending on the proposal number *N*) or not, each acceptor node will send a promise packet signifying the agreement to process the request. If the proposer node receives promise packets by majority of the acceptor nodes, it proceeds with the algorithm else it tries to reinitiate the proposal with *N* as the highest value *N'* received from the acceptor nodes. If out of the six acceptor nodes one fails and the proposer node is unaware of the failure then even if three of the acceptor nodes reply with promise packets in response to a prepare packet (a majority), the proposer node will try to reinitiate the proposal since according to the proposer node the total number of acceptor nodes is six whereas the number of acceptor nodes in reality is five.

The proposer node along with the acceptor nodes which are a part of the group, keep a watch on the group for acceptor node failures. Upon a failure, the proposer node is notified of the acceptor node failure. Once the failure is detected the proposer node updates its count of total acceptor nodes in the system and performs other book-keeping tasks specific to the system. In case, a proposer node fails, first the leader election is performed and upon election the newly elected acceptor node performs the book-keeping tasks.

## 6. Confidentiality and Integrity of system specific packets

Confidentiality and integrity of the system specific data packets mentioned in section 2is important for the proper functioning of the suggested system. A scenario in which system specific packet integrity violation can affect the system functioning is as follows: if an unauthorized third party sniffs a prepare packet initiated by the proposer node and attempts to simulate a promise packets in response to the proposer node then even if majority of the actual acceptor nodes do not reply back with the promise packet, the proposer node will initiate the accept request packet, which could affect the further functioning of the system.

Confidentiality and integrity of the system specific packets exchanged between the system components can be ensured by transforming and fragmenting the data packets before transmitting them as suggested in [12]. [5] suggests a similar approach towards secure data transfer over a network using the concept of *jigsaw puzzle*. Another approach based on the LSB data hiding technique suggested in [8] [9] can be used. [4] discusses different steganography techniques that can be applied for ensuring confidentiality and integrity of the system specific data packets. In case of the deployment of the suggested system in an ad-hoc mobile network, routing techniques mentioned in [11] can be used for propagating the system specific packets among the system components.

## 7. Conclusion

In conclusion, the suggested state machine replication approach based on the PAXOS algorithm is instrumental in detecting abnormalities in the system. Detection of abnormality in the system based on the state machine replication approach ensures accuracy, as the abnormality is detected only if no agreement is achieved among the replicated instances of the system.